\documentclass[10pt,aps,reprint,prb,floatfix,superscriptaddress,longbibliography]{revtex4-2}
\pdfoutput=1
\usepackage{graphicx,latexsym}
\usepackage{dcolumn,amsfonts}
\usepackage{amssymb,amsmath,bm}

\usepackage[breaklinks=true,pdfencoding=auto]{hyperref}

\usepackage{natbib}
\usepackage{balance}

\hypersetup{
	colorlinks   = true, 
	urlcolor     = blue, 
	linkcolor    = blue, 
	citecolor    = red 
}

\usepackage{color}
\usepackage{ulem}

\begin{document}

	\title{The tuning of para- and diamagnetic cavity photon excitations\\ in a square array
	       of quantum dots in a magnetic field}

\author{Vidar Gudmundsson}
	\email{vidar@hi.is}
	\affiliation{Science Institute, University of Iceland, Dunhaga 3, IS-107 Reykjavik, Iceland}
	\author{Vram Mughnetsyan}
	\email{vram@ysu.am}
	\affiliation{Department of Solid State Physics, Yerevan State University, Alex Manoogian 1, 0025 Yerevan, Armenia}
	\author{Hsi-Sheng Goan}
	\email{goan@phys.ntu.edu.tw}
    \affiliation{Department of Physics and Center for Theoretical Physics, National Taiwan University, Taipei 106319, Taiwan}
	\affiliation{Center for Quantum Science and Engineering, National Taiwan University, Taipei 106319, Taiwan}
	\affiliation{Physics Division, National Center for Theoretical Sciences, Taipei 106319, Taiwan}
	\author{Jeng-Da Chai}
	\email{jdchai@phys.ntu.edu.tw}
	\affiliation{Department of Physics and Center for Theoretical Physics, National Taiwan University, Taipei 106319, Taiwan}
	\affiliation{Center for Quantum Science and Engineering, National Taiwan University, Taipei 106319, Taiwan}
	\affiliation{Physics Division, National Center for Theoretical Sciences, Taipei 106319, Taiwan}
	\author{Nzar Rauf Abdullah}
	\email{nzar.r.abdullah@gmail.com}
	\affiliation{Physics Department, College of Science,
		University of Sulaimani, Kurdistan Region, Iraq}
	\author{Chi-Shung Tang}
	\email{cstang@nuu.edu.tw}
	\affiliation{Department of Mechanical Engineering, National United University, Miaoli 36003, Taiwan}
	\author{Valeriu Moldoveanu}
	\email{valim@infim.ro}
	\affiliation{National Institute of Materials Physics, PO Box MG-7, Bucharest-Magurele,
		Romania}
	\author{Andrei Manolescu}
	\email{manoles@ru.is}
	\affiliation{Department of Engineering, Reykjavik University, Menntavegur
		1, IS-102 Reykjavik, Iceland}

%

\begin{abstract}
We employ a ``real-time'' excitation scheme to calculate the excitation spectra of
a two-dimensional electron system in a square array of quantum dots placed in a circular cylindrical far-infrared photon cavity subjected to a perpendicular homogeneous external
magnetic field. The Coulomb interaction of the electrons is handled via spin density
functional theory and the para- and the diamagnetic parts of the electron-photon coupling
are updated according to a configuration interaction method in each iteration of
the density functional calculation.
The results show that an excitation scheme built on using the symmetry of the
lateral square superlattice of the dots and the cylindrical cavity produces both para- and
diamagnetic resonance peaks with oscillator strengths that can be steered by the excitation
pulse parameters. The excitation method breaks the conditions for the generalized Kohn
theorem and allows for insight into the subband structure of the electron system and can
be used both in and outside the linear response regime.

\end{abstract}

\maketitle
%
%

\section{Introduction}
Far-infrared (FIR) spectroscopy of various types has the last four decades been
important to study carrier excitations in the conduction band of two-dimensional
electron systems, homogeneous or modulated. The extraordinary purity and mobility
of the two-dimensional electron gas (2DEG) in GaAs and related heterostructures has been
instrumental in establishing bounds on theoretical models and aid in their development
\cite{Ando82:437,Batke86:6951,PhysRevB.38.12732,PhysRevB.44.9122,PhysRevB.61.R16319}.

Early on, it was realized that FIR spectroscopy has its limitations set
by general symmetry arguments, most notable being the original Kohn theorem stating that
the cyclotron resonance is not influenced by the Coulomb interaction as long as
the exciting electrical field can be viewed as a homogeneous rotating microwave
field and the kinetic term for the electrons is quadratic in the momentum
\cite{Kohn61:1242}.
Later on, this was expanded to the so-called generalized Kohn theorem stating that
the FIR excitation of parabolically confined electrons in quantum dots and wires
can only excite their center of mass (CM) motion if the wavelength of the exciting
electrical field is much longer than the characterized length of the electronic
system \cite{Bakshi90:7416,Maksym90:108}.
Even though the Kohn theorems tend to ``block the view'' of the internal dynamics
of the particular electron systems, they play an important role as tests on the
accuracy of analytical and numerical models.
Moreover, their prerequisites are not always totally fulfilled for
experimental systems, so many deviations from them are known. For example, the confinement is not
always parabolic \cite{Gudmundsson91:12098,Gudmundsson95:17744,Bollweg96:2774,Darnhofer96:591},
with Raman scattering a finite wavevector
is imposed on the system \cite{DAHL1994441,Steinebach99:10240}, the kinetic energy
in GaAs can be slightly nonparabolic for higher order processes \cite{Maag2016},
and in graphene and graphene-like systems the kinetic energy is linear in the momentum operator
\cite{PhysRevLett.104.067404}.

The successful placement of extended 2DEGs in FIR photon cavities
and external homogeneous magnetic field opens up further
avenues to explore their properties and has instigated researchers to examine
the regime of nonperturbative electron-light coupling
\cite{doi:10.1126/science.1258595,Zhang1005:2016}.
Concurrently, to the experimental development, several directions have been taken
in the theoretical modeling of extended or localized electron systems in
microcavities \cite{PhysRevA.90.012508,doi:10.1073/pnas.1518224112},
or atomic systems in chemistry \cite{10.1063/1.5142502,flick2021simple,Flick2017},
under the general name of quantum electrodynamical
density functional theory (QEDFT). Here, we will follow the general outlines
of Malave {\it et.\ al.}\ under the acronym of QED-DFT-TP, where TP stands for the
tensor product of the electron and the photon states used in their approach
\cite{10.1063/5.0123909}.

Parallel to this development, research groups are working on cavity quantum materials
of various types \cite{10.1063/5.0083825}.
In this work we present numerical calculations which show how a specially designed FIR-photon cavity and a time-dependent excitation can be used to shed light on the internal dynamics of
the electrons in the 2DEG-cavity system, and how it is possible to select the type of
the underlying processes determining the characteristics of an excited state of the system.

In continuation of the previous modeling of the static state of a 2DEG in an array of quantum
dots in an external magnetic field placed in a cylindrical FIR-cavity
\cite{PhysRevB.109.235306}, we show how time-dependent excitation of the system
can be implemented in order to emphasize magnetically active photon processes controlling
the ratio of virtual and real para- and diamagnetic transitions to reach states beyond
the traditional linear response regime \cite{Kubo57:570}.

The paper is organized as follows: In Sec.\ \ref{Model} we describe
the model. The results and discussion thereof are found in
Sec.\ \ref{Results}, with the conclusions drawn in Sec.\ \ref{Conclusions}.

\section{Model}
\label{Model}
The 2DEG considered in this work is in a lateral square superlattice of quantum dots in a
GaAs heterostructure subjected to a perpendicular homogeneous external magnetic field.
The effective mass of the electrons is $m^*=0.067m_e$,
the dielectric constant $\kappa = 12.4$, and the effective $g$-factor is
$g^* = -0.44$.

\subsection{The static system}
\label{Static-system}

The Hamiltonian of the static 2DEG-cavity system is
\begin{equation}
    H = H_\mathrm{e} + H_\mathrm{int} + H_\gamma,
\label{Hinitial}
\end{equation}
where
\begin{equation}
    H_\mathrm{e} = H_0 + H_\mathrm{Zee} + V_\mathrm{H} + V_\mathrm{per} + V_\mathrm{xc},
\label{He}
\end{equation}
describes the 2DEG in a square array of quantum dots and
\begin{equation}
    H_0 = \frac{1}{2m^*}\bm{\pi}^2, \quad\mbox{with}\quad
    \bm{\pi} = \left(\bm{p}+\frac{e}{c}\bm{A} \right).
\label{H0}
\end{equation}
The external vector potential ${\bm A} = (B/2)(-y,x)$ generates the homogeneous
magnetic field $\bm{B}=B\hat{\bm{z}}$ perpendicular to the $xy$-plane of the 2DEG.
The spin Zeeman term is $H_\mathrm{Zee} = \pm g^* \mu_\textrm{B}^* B/2$, with
$\mu_\textrm{B}^*$ the effective Bohr magneton.
The direct Coulomb interaction in terms of the effective local charge density
$\Delta n(\bm{r}) = n_\mathrm{e}(\bm{r})-n_\mathrm{b}$ is
\begin{equation}
    V_\mathrm{H}(\bm{r}) = \frac{e^2}{\kappa}\int_{\mathbf{R}^2}d\bm{r}'\frac{\Delta n(\bm{r}')}
    {|\bm{r}-\bm{r}'|},
\label{Vcoul}
\end{equation}
where $-en_\mathrm{e}(\bm{r})$ is the electron charge density and
$+en_\mathrm{b}$ is the homogeneous positive background charge density
representing the ionized donors guaranteeing the charge neutrality of the total system.
The square array of quantum dots is represented by the periodic potential
\begin{equation}
    V_\mathrm{per}(\bm{r}) = -V_0\left[\sin \left(\frac{g_1x}{2} \right)
    \sin\left(\frac{g_2y}{2}\right) \right]^2,
\label{Vper}
\end{equation}
where $V_0 = 16.0$ meV. $V_\mathrm{per}(\bm{r})$ is depicted in Fig.\ 1 in Ref.\
\cite{PhysRevB.108.115306}.
The superlattice is spanned by the spatial vectors
$\bm{R}=n\bm{l}_1+m\bm{l}_2$ with $n,m\in \bm{Z}$,
where the unit vectors are $\bm{l}_1 = L\bm{e}_x$ and $\bm{l}_2 = L\bm{e}_y$,
and the reciprocal lattice is spanned by $\bm{G} = G_1\bm{g}_1 + G_2\bm{g}_2$ with
$G_1, G_2\in \mathbf{Z}$ and the unit vectors $\bm{g}_1 = 2\pi\bm{e}_x/L$
and $\bm{g}_2 = 2\pi\bm{e}_y/L$.
The period of the superlattice is $L = 100$ nm. The Coulomb exchange and correlation potentials $V_\mathrm{xc}$ of the local spin density approximation (LSDA) are documented in Appendix A of
Ref.\ \cite{PhysRevB.106.115308}. The interaction of the 2DEG with the quantized vector potential,
${\bm A}_\gamma$, of the photon cavity in terms of the electron current, and charge densities is
\begin{align}
    H_\mathrm{int} = \frac{1}{c}\int_{\mathbf{R}^2} d\bm{r}\; &
    {\bm J}({\bm r})\cdot{\bm A}_\gamma (\bm{r}) \nonumber\\
    +& \frac{e^2}{2m^*c}\int_{\mathbf{R}^2} d\bm{r}\;
    n_\mathrm{e}(\bm{r})A^2_\gamma(\bm{r}).
\label{e-g}
\end{align}
The two terms of the interaction are the para- and the diamagnetic interactions, respectively.
In the Appendix of Ref.\ \cite{PhysRevB.109.235306} the electron-photon coupling (\ref{e-g})
for a single quantized TE$_{011}$ mode of a cylindrical cavity is derived in the long wavelength approximation, i.e.\ when the spatial variation of the far-infrared cavity field is small
on the scale of $L$.
In terms of the creation and annihilation operators of the eigenstates $|n\rangle$ of
the photon number operator, $a^\dagger_\gamma a_\gamma$, the interaction is
\begin{align}
    H_\mathrm{int} &= g_\gamma \hbar\omega_c \left\{ lI_x + lI_y\right\} \left(a^\dagger_\gamma + a_\gamma\right)\nonumber\\
    &+ g^2_\gamma \hbar\omega_c {\cal N}\left\{\left(a^\dagger_\gamma a_\gamma + \frac{1}{2}\right)
    +\frac{1}{2}\left(a^\dagger_\gamma a^\dagger_\gamma + a_\gamma a_\gamma\right)\right\}
\label{e-gIxIyN}
\end{align}
with the integrals, $I_x$, $I_y$, and ${\cal N}$, that can be interpreted as functionals
of the current and the charge densities, respectively, defined in the Appendix of
Ref.\ \cite{PhysRevB.109.235306}. The functional interpretation will be essential in
understanding the subsequent computational approach.
$\omega_c = eB/(m^*c)$ is the cyclotron frequency.
It is possible to define a dimensionless coupling strength as
\begin{equation}
    g_\gamma = \left\{ \left( \frac{e{\cal A}_\gamma}{c} \right) \frac{l}{\hbar} \right\},
\end{equation}
where $l = (\hbar c/(eB))^{1/2}$ is the magnetic length, that sets a natural length scale
competing with the superlattice length $L$.
The fundamental energy of the photon mode is $E_\gamma = \hbar\omega_\gamma$ and the free photon
Hamiltonian is
\begin{equation}
    H_\gamma = \hbar\omega_\gamma a^\dagger_\gamma a_\gamma
\end{equation}
with the zero point energy of the photon mode neglected.
In the Appendix of Ref.\ \cite{PhysRevB.109.235306}
the vector potential of the cylindrical TE$_{011}$ cavity photon mode in the long wavelength approximation is derived as
\begin{equation}
    {\bm A}_\gamma (\bm{r}) = \bm{e}_\phi {\cal A}_\gamma\left(a^\dagger_\gamma + a_\gamma  \right)
    \left(\frac{r}{l} \right)
\end{equation}
with $\bm{e}_\phi$ the unit angular vector in the polar coordinates.
This vector potential has the same spatial form as the vector potential ${\bm A}$ generating the
external homogeneous magnetic field $\bm{ B} = B\bm{e}_z$ \cite{PhysRevB.109.235306}.

We employ a quantum electrodynamical density functional theory approach, QED-DFT-TP initially presented by Malave
\cite{10.1063/5.0123909} and recently adapted to our 2DEG-cavity system by calculating
the energy spectrum and the eigenstates of $H$ (\ref{Hinitial}) in a linear functional basis constructed as a tensor product (TP) of the electron and the photon states
\begin{equation}
    |\bm{\alpha\theta}\sigma n\rangle = |\bm{\alpha\theta}\sigma\rangle\otimes|n\rangle.
\label{TP}
\end{equation}
The photon states $|n\rangle$ are, as stated above, the eigenstates of the photon number operator, and the electron states $|\bm{\alpha\theta}\sigma \rangle$ are the single-electron states
of Ferrari constructed
for the periodic 2DEG in an external magnetic field at each point in the first Brillouin zone, i.e.\ $\bm{\theta} = (\theta_1,\theta_2)\in [-\pi,\pi]\times[-\pi,\pi])$
\cite{Ferrari90:4598,Silberbauer92:7355,Gudmundsson95:16744,PhysRevB.105.155302,PhysRevB.106.115308}, and $\sigma\in\{\uparrow,\downarrow\}$ is the quantum number for the $z$-component of the electron spin.
All the other quantum numbers of the Ferrari states \cite{Ferrari90:4598} are combined in $\bm{\alpha}$, which can be interpreted as a subband index.

The single-electron states of Ferrari satisfy the commensurability condition for the aforementioned competing length scales in the system, that can be expressed as
$B{\cal A} = BL^2 = pq\Phi_0$ in terms of the quantum of the magnetic flux, $\Phi_0 = hc/e$,
and the integers $p$ and $q$
\cite{Hofstadter76:2239,Ferrari90:4598,Silberbauer92:7355,Gudmundsson95:16744}.
Each Landau-band in the energy spectrum is split into $pq$ subbands.
It is possible to express the commensurability condition in different ways \cite{Hofstadter76:2239,Ferrari90:4598,PhysRevB.106.085140},
but it reflects the fact that spatial translations by superlattice vectors in the external magnetic field gather a Peierls phase and have to be replaced by magnetotranslations.
In the Appendix of Ref.\ \cite{PhysRevB.109.235306}, it is shown that the electron-photon
coupling (\ref{e-g}) and the vector potential $\bm{A}_\gamma$ do not break the spatial symmetry
of the 2DEG system in the external magnetic field and the periodic quantum dot potential.

\subsection{The dynamic system}
\label{Dynamic-system}
The real-time excitation of the 2DEG-cavity system is accomplished by applying a
short time-dependent modulation of the electron-photon coupling described by a
time-dependent Hamiltonian $H_\mathrm{ext}$, in addition to the total Hamiltonian
(\ref{Hinitial}),
\begin{align}
    H_\mathrm{ext}(t) &= F(t)
    \biggl[ g_\gamma \hbar\omega_c\left\{lI_x + lI_y \right\} \left( a^\dagger + a \right)\nonumber\\
    + &\left. g_\gamma^2 \hbar\omega_c {\cal N} \left\{ \left( a^\dagger a + \frac{1}{2}\right)
    + \frac{1}{2}\left( a^\dagger a^\dagger + a a  \right) \right\} \right]
\label{Ht}
\end{align}
with
\begin{equation}
    F(t) = \left( \frac{V_t}{\hbar\omega_c}\right) (\Gamma t)^2 \exp{(-\Gamma t)}\cos{(\omega_\mathrm{ext} t)},
\label{ft}
\end{equation}
where $\omega_\mathrm{ext}$ is the frequency of the modulation of the electron-photon
interaction. Both, the para- and the diamagnetic interactions become time-dependent.
This excitation scheme does not lead to direct mixing of states at different
$\bm{\theta}$-points in the Brillouin zone and enables the use of the Liouville-von Neumann
equation for the time-evolution of both spin components of the
density operator $\rho^{\bm{\theta}}$ at each $\bm{\theta}$
\begin{equation}
    i\hbar\partial_t \rho^{\bm{\theta}} (t) = \left[ H[\rho^{\bm{\theta}}(t)], \rho^{\bm{\theta}}(t)\right]
\label{L-vN}
\end{equation}
implying that the time-dependent Hamiltonian is a functional of the density operator
$\rho^{\bm{\theta}}$ through the electron charge and current densities, that have to be
updated in each iteration within the time-steps of the numerical time integration of
Eq.\ (\ref{L-vN}). We use the Crank-Nicolson scheme for this as it is advantageous
for Hermitian systems \cite{Crank1947}.
Note that the excitation scheme is in a quantized form regarding the
cavity photons and includes all the fundamental photon processes inherent in the
electron-photon interaction.
Moreover, as the interactions and the excitation do not lead to nonvanishing
matrix elements between different $\bm{\theta}$ in the reciprocal space, we can
extend the dynamical calculation to a larger state space than was possible
for the system in the Hartree or the QEDFT approach where the excitation had
a finite wavevector \cite{PhysRevB.105.155302,PhysRevB.108.115306}.
Here, we have thus a larger contribution from higher order electron-photon processes
of various types.

The initial conditions for the density operators at $t = 0$ can
be set in the linear basis that results from the diagonalization
procedure for the static case $\{|\bm{\alpha\theta}\sigma)\}$ with the associated
wavefunctions $\psi_{\bm{\alpha\theta}\sigma}(\bm{r}) = \langle\bm{r}|\bm{\alpha\theta}\sigma)$,
while the states $|\bm{\alpha\theta}\sigma\rangle$, the Ferrari states, correspond
to the wavefunctions $\phi_{\bm{\alpha\theta}\sigma}(\bm{r}) =
\langle\bm{r}|\bm{\alpha\theta}\sigma\rangle$.
Note specially, the notation used with different brackets:
$|\bm{\alpha\theta}\sigma \rangle$ for the Ferrari states,
and $|\bm{\alpha\theta}\sigma )$ for the Coulomb interacting cavity-photon dressed
electron states.
If in the $\{|\bm{\alpha\theta}\sigma)\}$-basis the initial conditions for the
corresponding density operators are the static equilibrium states
\begin{equation}
    r^{\bm{\theta}}_{\bm{\alpha}\sigma,{\bm{\beta}}\sigma'}(0) = f\left(E_{\bm{\alpha\theta}\sigma} - \mu \right)
    \delta_{\bm{\alpha},\bm{\beta}}\delta_{\sigma,\sigma'},
\label{rho_0_int}
\end{equation}
where $f$ denotes the equilibrium Fermi distribution,
then the initial conditions in the $\{|\bm{\alpha\theta}\sigma\rangle\}$ can be expressed as
\begin{equation}
    \rho^{\bm{\theta}}(0) = W^{\bm{\theta}}\left\{ r^{\bm{\theta}}(0)\right\} (W^{\bm{\theta}})^\dagger
\label{rho_0}
\end{equation}
implying that the Liouville-von Neumann equation (\ref{L-vN}) is solved in the
same functional tensor product basis (\ref{TP}) as the Hamiltonian for the static
system is diagonalized in, delivering its energy spectrum $E_{\bm{\alpha\theta}\sigma}$
and chemical potential $\mu$. $W^{\bm{\theta}}$ is the unitary transformation between
the two bases in each point $\bm{\theta}$ of the first Brillouin zone.
The initial density matrix is thus not diagonal for computational convenience and,
additionally, the initial excitation promotes off-diagonal elements.
The time dependent electron density is calculated as
\begin{align}
    n_\sigma (\bm{r},t)& =\nonumber\\
    \frac{1}{(2\pi)^2}&\int_{-\pi}^\pi d\bm{\theta}\,
    \sum_{\bm{\alpha\beta}}\phi^*_{\bm{\alpha\theta}\sigma}(\bm{r})
    \phi_{\bm{\beta\theta}\sigma}(\bm{r})\rho^{{\bm{\theta}}}_{\bm{\alpha}\sigma,\bm{\beta}\sigma}(t)
\label{Net}
\end{align}
and the time-dependent average photon number
\begin{equation}
     N_\gamma (t) = \frac{1}{(2\pi)^2}\sum_{\sigma}\int^{\pi}_{-\pi} d\bm{\theta}\; \mathrm{Tr} \left\{ \rho^{\bm{\theta}}_{\sigma}(t) a^\dagger_\gamma a_\gamma\right\}
\label{Ngt}
\end{equation}
is conveniently evaluated in the $\{|\bm{\alpha\theta}\sigma\rangle\}$-basis.
The external magnetic field leads to rotational currents in the system which the
excitation pulse modulates, so we calculate the dynamical orbital magnetization
to monitor the current oscillations \cite{Desbois98:727,Gudmundsson00:4835}
\begin{equation}
    Q_J(t) = \frac{1}{2c{\cal A}}\int_{\cal A} d\bm{r} \left( {\bf r}\times
    \langle {\bf J}({\bf r},t) \rangle \right) \cdot{\bm{e}_z},
    \label{Mo}
\end{equation}
with ${\cal A} = L^2$. The time-dependent mean charge current density in the system is
\begin{align}
    \bm{J}_{i}(\bm{r},t) = \frac{-e}{m^*(2\pi)^2}\sum_{\bm{\alpha\beta}\sigma}\int_{-\pi}^{\pi} d\bm{\theta}\;
    \Re&\left\{ \phi_{\bm{\alpha\theta}\sigma}^*(\bm{r})\bm{\pi}_i \phi_{\bm{\beta\theta}\sigma}(\bm{r}) \right\}\nonumber\\
    &\rho^{\bm{\theta}}_{\bm{\beta\theta}\sigma,\bm{\alpha\theta}\sigma}(t)
    \label{currD}
\end{align}
for the Cartesian components $i=x$ or $y$. The label $J$ in $Q_J$ is not an integer,
but refers to the charge current. As the excitation does not break the initial
square symmetry of the superlattice the current and the electron densities retain that
periodicity. The spatial operators $\bm{x}$ and $\bm{y}$ are not periodic so we calculate
their mean values and those of their combinations from the dynamical electronic
density (\ref{Net}).
Even though the mean values of the spatial operators are not shown here, we need them to monitor
that the center of mass of the electronic system remains stationary with high accuracy, and in
order to monitor the small monopole (breathing), $Q_0$, and quadrupole oscillations, $Q_2$,
of the electron density promoted by the excitation, the superlattice square symmetry, and the Lorentz force \cite{PhysRevB.105.155302,PhysRevB.108.115306}.
The total energy of the photon-dressed 2DEG is calculated as
\begin{align}
\label{Etott}
      E_\mathrm{total} =& \mathrm{Tr}\left\{\left( H[\rho (t)] + H_\mathrm{ext}[\rho (t); t]\right)\rho(t)\right\}\nonumber\\
      -& \frac{1}{2}\int_{\bm{R}^2} d{\bm{r}}d{\bm{r}'}\; \frac{e^2\Delta n(\bm{r},t)\Delta n(\bm{r}',t)}{\kappa|\bm{r}-\bm{r}'|}\\
      -& \sum_\sigma \int_{\bm{{\cal A}}} d{\bm{r}}\;  n_\sigma (\bm{r},t) V_\mathrm{xc,\sigma}(\bm{r},t)
      + E_\mathrm{xc,\sigma}[n_\sigma (\bm{r},t)], \nonumber
\end{align}
where $n_\uparrow + n_\downarrow = n_\mathrm{e}$.
$E_\mathrm{xc,\sigma}$ are the Coulomb exchange correlation functionals, and
$V_\mathrm{xc,\sigma}$ the corresponding potentials (see Appendix A in
Ref.\ \onlinecite{PhysRevB.106.115308}). The second line in Eq.\ (\ref{Etott})
represents a counter term to the double counting of the Coulomb interaction in the first line
of the equation. It is common to mean-field and DFT models.

\section{Results}
\label{Results}
In order to build up the understanding of the excitation of the system
we start with one electron in each quantum dot of the square superlattice
in a cylindrical cavity interacting with the photons of the single TE$_{011}$
mode. Even though only one electron resides in each dot it interacts in the sense of DFT
through the charge density with the electrons in the other dots,
and the exchange Coulomb interaction leads to a large spin splitting in the finite
magnetic field due to the unpaired spin in each dot \cite{PhysRevB.109.235306}.

\subsection{One electron in a dot, $N_\mathrm{e}=1$, variation of the
            photon energy $E_\gamma$}
\label{varEg}
\begin{figure}[htb]
    \hspace*{0.5cm}
    \vspace*{-0.6cm}
    \includegraphics[width=0.38\textwidth]{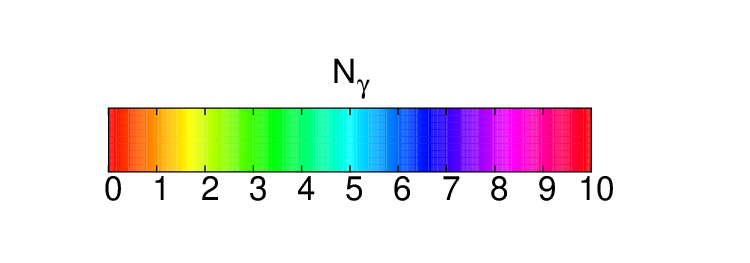}
    \includegraphics[width=0.42\textwidth]{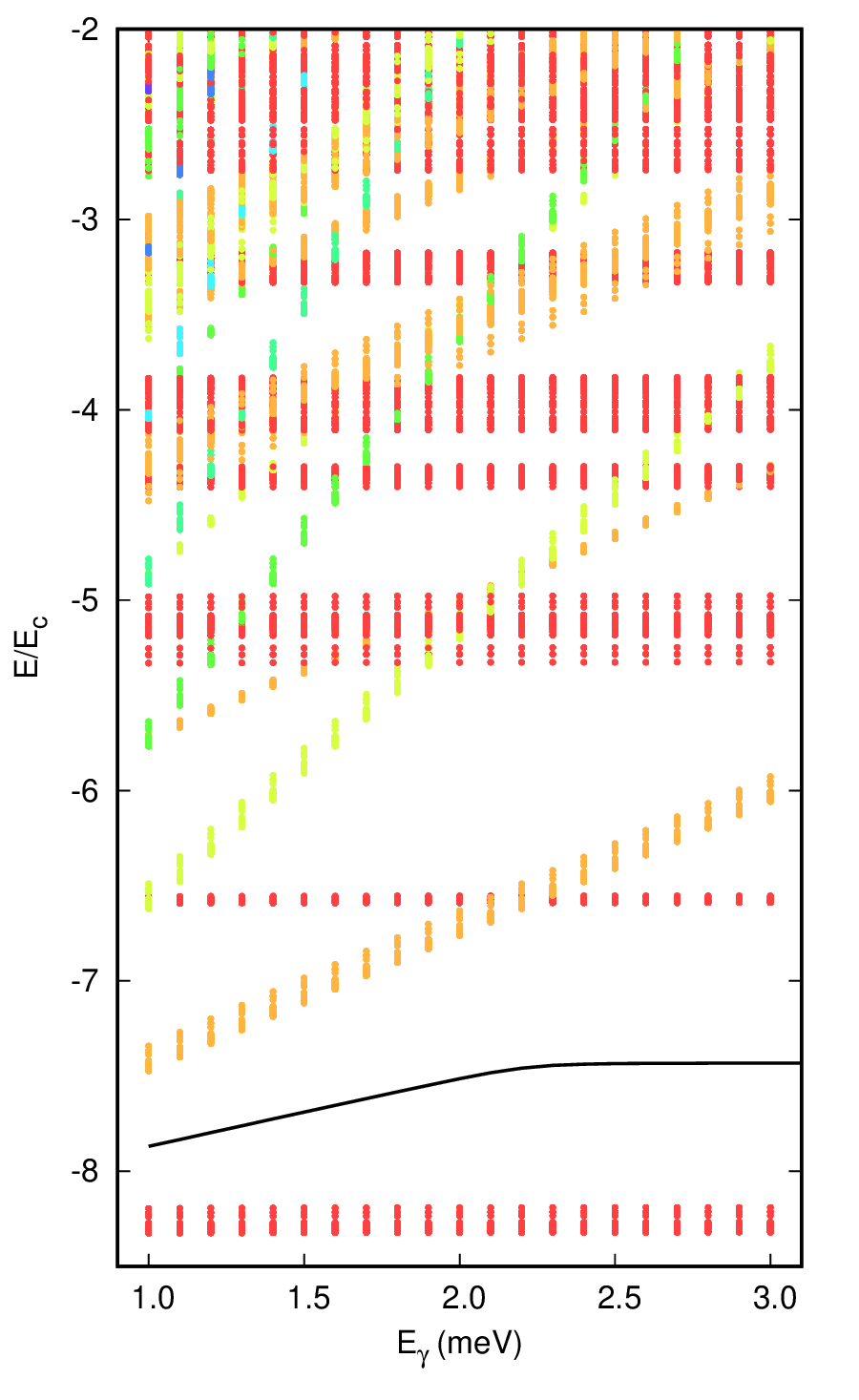}
\caption{The width of the energy subbands as a function of the photon
         energy $E_\gamma$. The photon content of the subbands is
         encoded in the color with red for 0 and violet for 10 photons
         as the colorbar at the top indicates.
         The black curve represents the chemical potential $\mu$.
         $N_\mathrm{e} = 1$, $pq = 2$ corresponding to $B = 0.8271$ T or
         $E_c = \hbar\omega_c = 1.4291$ meV, $g_\gamma = 0.08$,
         $T = 1.0$ K, and $L= 100$ nm.}
\label{E_Eg}
\end{figure}

We start by exploring the effects of the photon energy $E_\gamma$
on the energy subbands, total energy and mean photon number $N_{\gamma}$.
We are not implying that this variation is easy to achieve in experiments, but it gives us valuable
insight into the features of the model. The static energy spectrum, showing the width of
the subbands, versus $E_\gamma$ is shown in Fig.\ \ref{E_Eg} with the mean photon number
color coded (as is indicated with the colorbar). The lowest subband, below the
black curve indicating the chemical potential $\mu$,
has a red color signifying a low photon component, i.e.\ the average number of photons in
a quantum dot is much closer to 0, than 1. The orange subband above $\mu$ is the first
photon replica of the ground state and its color indicates that its mean photon number is
close to 1. The third horizontal red narrow subband is the other spin component of the
ground state, not occupied. Clearly, the photon replicas have a ``slope'' that indicates
their mean photon number as each photon has the energy $E_\gamma$.
As the states in each subband are cavity photon-dressed electron states, the mean
photon number of a subband does not have to be an integer, especially for strong
electron-photon coupling.
The width of the lower subbands is small as an electron in them is rather localized in
each dot potential. The width of the unoccupied higher energy spin component of the
ground state subband is smallest as its ``DFT-interaction'' through the densities
is weak.

\begin{figure}[htb]
	\includegraphics[width=0.48\textwidth]{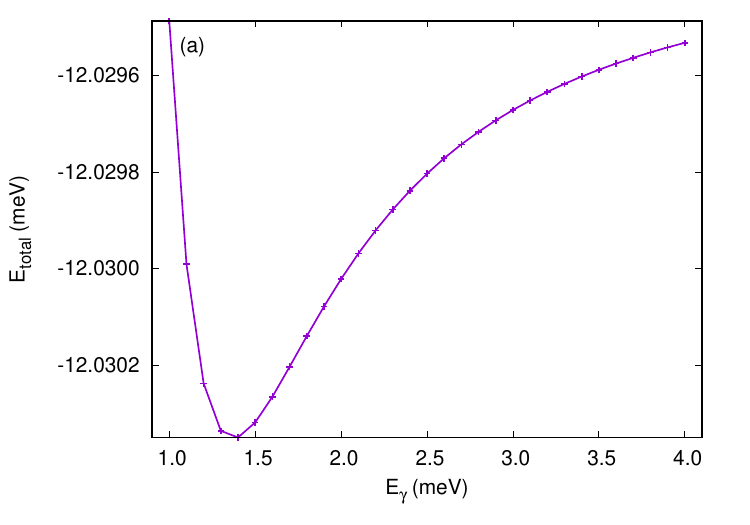}\\
	\includegraphics[width=0.48\textwidth]{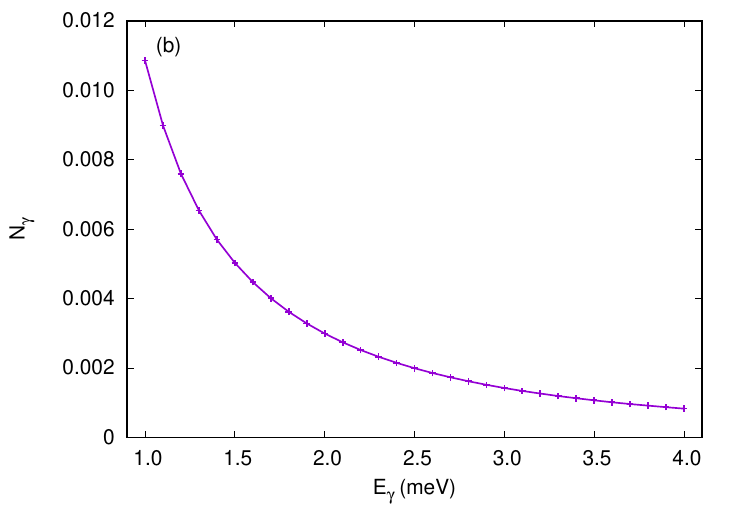}
\caption{The total energy  $E_\mathrm{total}$ (a), and the total mean
         photon number $N_\gamma$ (b)
	     of the system per unit cell as a function of the
         photon energy $E_\gamma$. $N_\mathrm{e} = 1$, $pq = 2$, $T = 1.0$ K, and $L= 100$ nm. }
\label{Etot-Ng}
\end{figure}

Fig.\ \ref{Etot-Ng} displays the total energy $E_\mathrm{total}$ (a) and
the mean photon number $N_\gamma$ (b) as functions of the photon energy $E_\gamma$.
The mean photon number decreases monotonously as the photon energy increases.
This is expected for $N_\mathrm{e}=1$, but can be very different for a higher number of electrons
in a quantum dot \cite{PhysRevB.109.235306}.

The total energy in Fig.\ \ref{Etot-Ng}(a) shows a very shallow
local minimum at 1.4 meV. Interestingly, this behavior is not found in a QEDFT
model with the functional used in \cite{PhysRevB.106.115308}. That functional
was adapted to a 2DEG in a homogeneous magnetic field from a functional originally
proposed by Flick \cite{flick2021simple} for 3D atomic systems. The functional,
a gradient based density functional for the electron-photon interactions, is
assembled with an approximation which includes all one-photon exchange
processes explicitly, while neglecting higher order processes. Of course the
self-consistent DFT iterations add effective higher order processes built on
these one-photon processes. The QED-DFT-TP approach \cite{10.1063/5.0123909} used
here includes multi-photon and higher order processes explicitly.
For $E_\gamma < E_c = \hbar\omega_c$
we have a similar monotonic decreasing of $E_\mathrm{total}$ with increasing
photon energy. Looking at the low part of the static energy spectrum in
Fig.\ \ref{E_Eg}, this is expected, but higher in the energy spectrum of the
2DEG in the quantum dots, the cyclotron energy $\hbar\omega_c$ slowly gains
an upper hand in the competition with the confinement energy of the dots.
When the photon energy surpasses the cyclotron energy, high order virtual processes
slightly add to the total energy. The shallow local minimum in the total energy is thus
a clear indication of the presence of high order virtual photon processes in
our model.

\begin{figure}[htb]
    \includegraphics[width=0.48\textwidth]{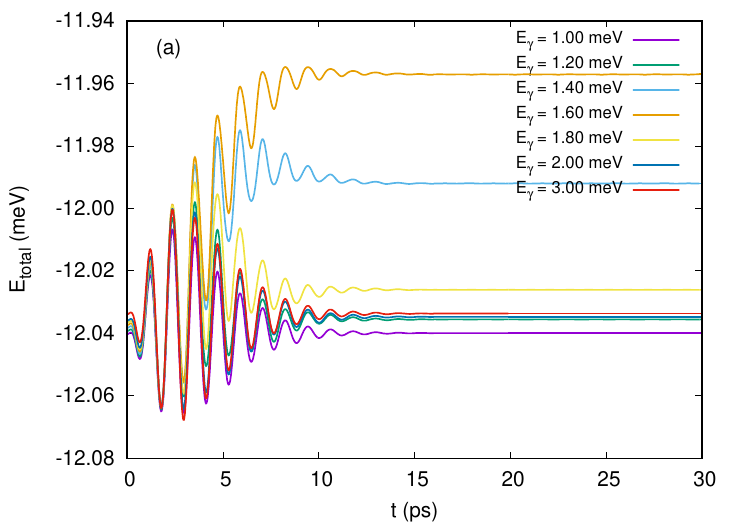}
    \includegraphics[width=0.48\textwidth]{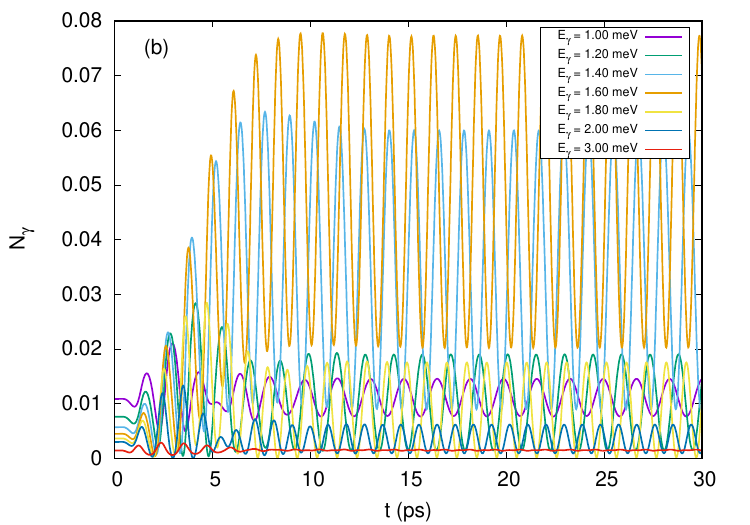}
\caption{The time evolution of the total energy $E_\mathrm{total}$ (a) and the total
         mean photon number $N_\gamma$ (b) for the first 30 ps for
         different values of the photon energy. $g_\gamma = 0.08$,
         $N_\mathrm{e} = 1$, $pq = 2$, $T = 1.0$ K, and $L= 100$ nm.
         $V_t/(\hbar\omega_c) = 0.8$, $\hbar\Gamma = 0.5$ meV,
         and $\hbar\omega_\mathrm{ext}=3.5$ meV.}
\label{Et-Ng-t}
\end{figure}

Now we turn our focus on the results of the ``real-time'' excitation of the
system. We use rather strong electron-photon interaction and strong excitation
with: $g_\gamma = 0.08$, $V_t/(\hbar\omega_c) = 0.8$, $\hbar\Gamma = 0.5$ meV,
and $\hbar\omega_\mathrm{ext}=3.5$ meV.
The initial state is the static equilibrium state determined by
Eq.\ (\ref{rho_0_int}).
The parameter choice will be analyzed below.
The time-evolution for the first 30 ps
of the total energy $E_\mathrm{total}$ and the mean photon number $N_\gamma$ are shown
in Fig.\ \ref{Et-Ng-t}.
The time series are calculated to $t = 100$ ps, but the excitation pulse is cut off
at $t = 16$ ps. The results are outside the regime of linear response
for a range of photon energy as energy is clearly pumped into the system.
After the excitation the mean total energy stays constant, but the
photon number oscillates steadily.
We analyze the Fourier power spectra for the time series of the
variables $N_\gamma$ and $Q_J$ for the time interval 16 - 100 ps.
We present the Fourier power spectra from $\omega/\omega_c = 0$ to 8, or
a lower number, without any smoothing or cutting off low frequency noise.
Since the system is excited slightly into the nonlinear regime we need to
analyze the role of the parameters of the excitation pulse.

\begin{figure}[htb]
    \includegraphics[width=0.48\textwidth]{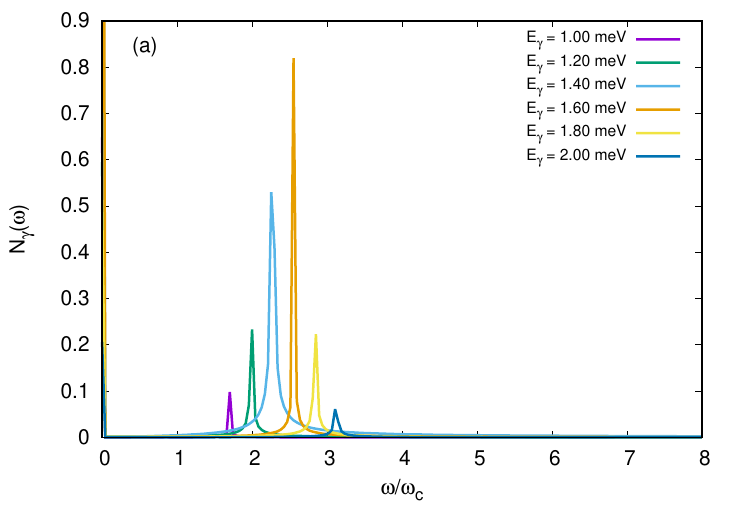}
    \includegraphics[width=0.48\textwidth]{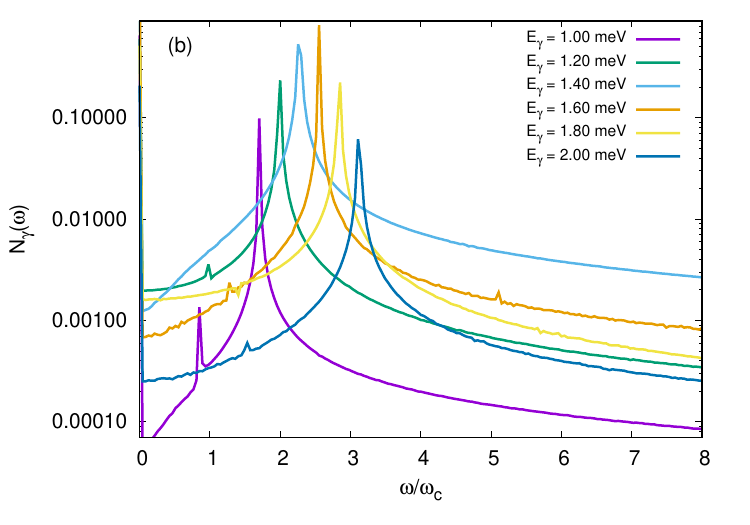}
\caption{The Fourier power spectra for the excitation of the total mean photon number
	     $N_\gamma$ for the system for
         several values of the photon energy $E_\gamma$ with a linear (a),
         or logarithmic scale (b). The Fourier transform is taken after
         the excitation pulse has vanished. $g_\gamma = 0.08$,
         $N_\mathrm{e} = 1$, $pq = 2$, $T = 1.0$ K, and $L= 100$ nm.
         $V_t/(\hbar\omega_c) = 0.8$, $\hbar\Gamma = 0.5$ meV,
         and $\hbar\omega_\mathrm{ext}=3.5$ meV.}
\label{Fp-Eg-Ng}
\end{figure}

The Fourier power spectra for the photon number $N_\gamma (\omega )$ is presented
in Fig.\ \ref{Fp-Eg-Ng} on a linear (a) and logarithmic scale (b)
for selected values of the photon energy $E_\gamma$.
The location of an excitation peak can be interpreted in terms of the energy
of the cavity photons and a shift caused by the electron-photon interactions.
The height of the peaks can be related to the parameters of the excitation pulse
which has a broad frequency range with a maximum around $\hbar\omega = 3.5$ meV
corresponding to $\omega/\omega_c\approx 2.5$, reflecting that
$\hbar\omega_\mathrm{ext} = 3.5$ meV.
At the end of this subsection the parameter
choice for the excitation pulse is addressed. In Fig.\ \ref{Fp-Eg-Ng} the excitation
peak for $E_\gamma = 1.4$ meV is just above $\omega/\omega_c = 2$, but a glance at
the static spectrum in Fig.\ \ref{E_Eg} shows the lowest photon replica with approximately
1 photon is barely over 1.4 meV above the ground state, corresponding to $\omega/\omega_c$ slightly
larger than 1. The excitation has thus mainly a contribution from a 2-photon diamagnetic transition, the second term in Eq.\ (\ref{e-g}), (\ref{e-gIxIyN}), and (\ref{Ht}).
The dispersion of the excitation peaks with increasing $E_\gamma$ confirms this idea as
the second photon replica in Fig.\ \ref{E_Eg} has a higher slope than the first
replica with respect to $E_\gamma$. Fig.\ \ref{Fp-Eg-Ng}(b) with the logarithmic
scale shows weak paramagnetic 1-photon peaks just around or above $\omega /\omega_c = 1$.
The diamagnetic e-photon interaction is much stronger than
the paramagnetic one in the cylindrical cavity with a TE$_{011}$ mode and a related
excitation pulse for the system in an external homogeneous magnetic field, but
we need to look closer at this.

\begin{figure}[htb]
    \includegraphics[width=0.48\textwidth]{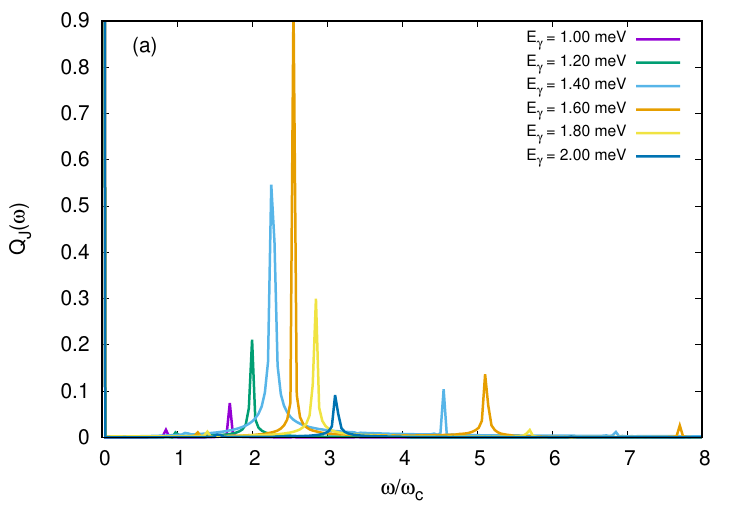}
    \includegraphics[width=0.48\textwidth]{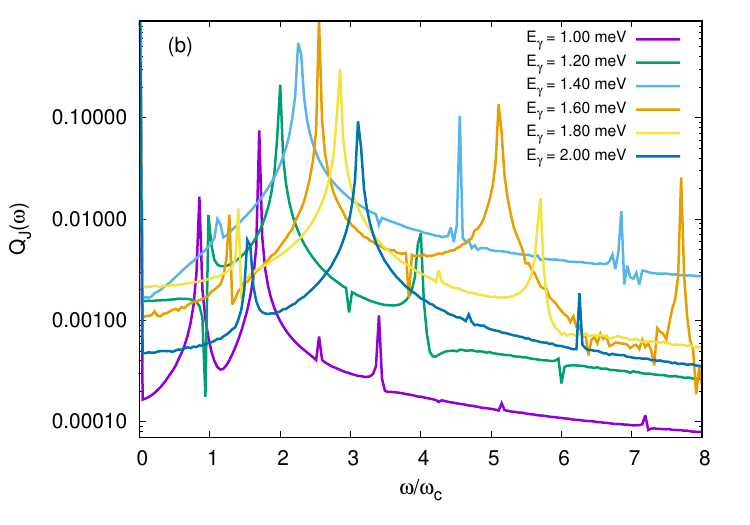}
\caption{The Fourier power spectra for the excitation of the mean orbital
	     magnetization $Q_J$ for the system for
	     several values of the photon energy $E_\gamma$ with a linear (a),
       	 or logarithmic scale (b). The Fourier transform is taken after
       	 the excitation pulse has vanished. $g_\gamma = 0.08$,
	     $N_\mathrm{e} = 1$, $pq = 2$, $T = 1.0$ K, and $L= 100$ nm.
	     $V_t/(\hbar\omega_c) = 0.8$, $\hbar\Gamma = 0.5$ meV,
	     and $\hbar\omega_\mathrm{ext}=3.5$ meV.}
\label{Fp-Eg-Qj}
\end{figure}

Further details emerge as the excitation power spectrum for the dynamic orbital
magnetization $Q_J(\omega )$ is viewed in Fig.\ \ref{Fp-Eg-Qj}.
We confirm that the location of the main peaks of $N_\gamma (\omega )$ and $Q_J(\omega )$
coincide, but $Q_J(\omega )$ displays several smaller peaks besides the main peak.
First, we recognize higher order diamagnetic transitions at energies a bit larger than
$4E_\gamma$ and $6E_\gamma$, but in addition there are also peaks at energies slightly larger
than $E_\gamma$, $3E_\gamma$, and $5E_\gamma$ that can be identified as paramagnetic
transitions.

\begin{figure}[htb]
	\includegraphics[width=0.48\textwidth]{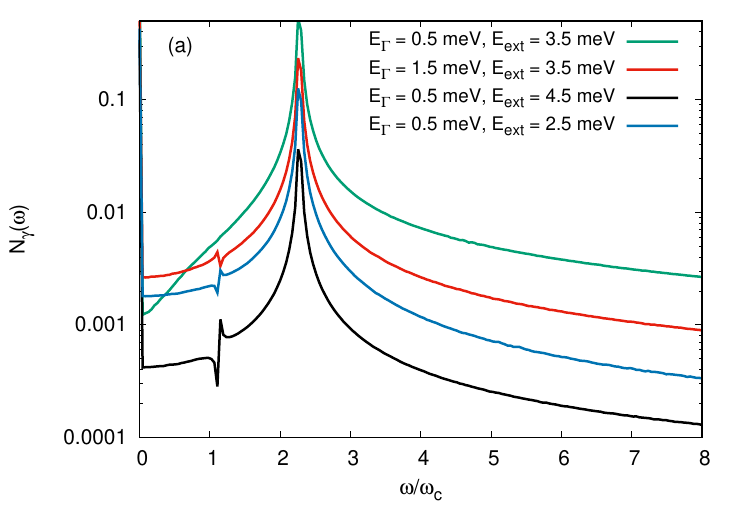}
	\includegraphics[width=0.48\textwidth]{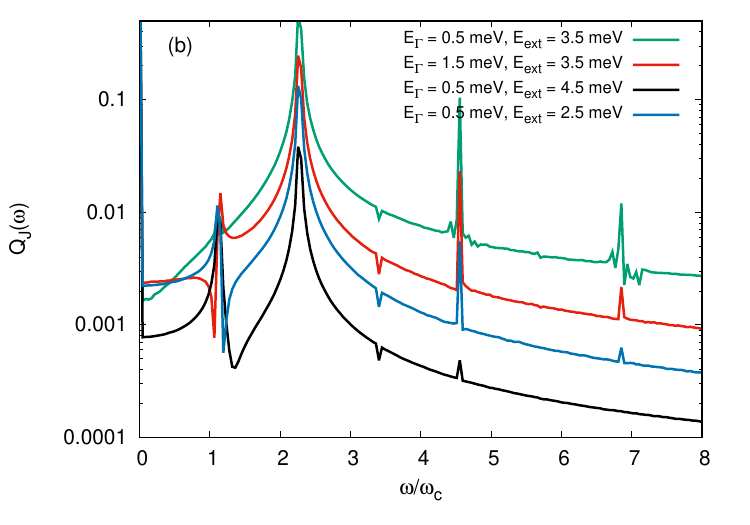}
\caption{The Fourier power spectra for the excitation of the
		 mean total photon number $N_\gamma$ (a) and the mean orbital
		 magnetization $Q_J$ (b) for the system for
		 $E_\gamma = 1.4$ meV. $E_\Gamma = \hbar\Gamma$,
		 and $E_\mathrm{ext} = \hbar\omega_\mathrm{ext}$.
		 The Fourier transform is taken after
		 the excitation pulse has vanished. $g_\gamma = 0.08$,
		 $N_\mathrm{e} = 1$, $pq = 2$, $T = 1.0$ K, and $L= 100$ nm.
		 $V_t/(\hbar\omega_c) = 0.8$.}
	\label{FP-varPulse}
\end{figure}

The excitation pulse induces time-dependent changes in the rotational current
distribution of the system, but for one electron in a dot the Lorentz force
only leads to small radial charge density (breathing) oscillations in the system. All the same,
the energy range offered by the excitation pulse with $\hbar\omega_\mathrm{ext} = 3.5$ meV
activates strong diamagnetic transitions, virtual and real.
In Fig.\ \ref{FP-varPulse}, one can see how a variation of the pulse frequency through
the parameter $\hbar\omega_\mathrm{ext}$, and its duration through
$\hbar\Gamma$, can influence the relative strength of the excitation peaks.
The shape of the excitation pulse can be used to tune the relative strength
of the pulses representing either mainly the fundamental para- or diamagnetic
transitions, or their respective higher harmonics.

\subsection{One electron in a dot, $N_\mathrm{e}=1$, variation of the
            excitation strength $V_t$}
\label{varVt}

\begin{figure}[htb]
    \includegraphics[width=0.48\textwidth]{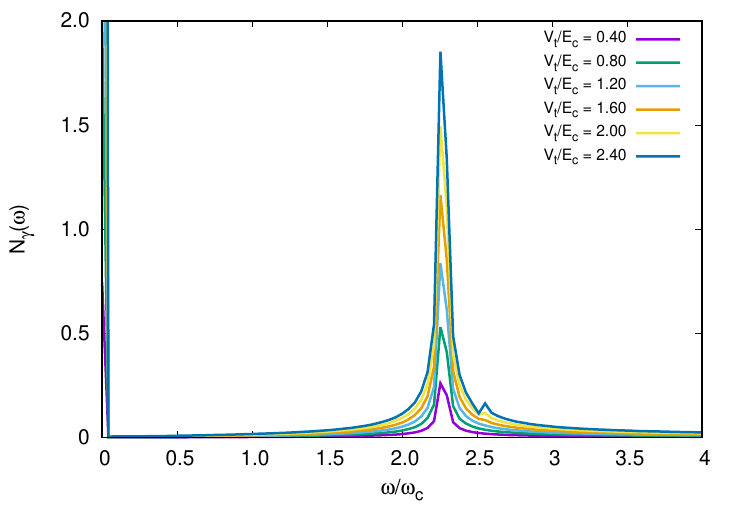}
\caption{The Fourier power spectra for the excitation of the total mean
	     photon number $N_\gamma$ for the system for
	     several values of the excitation strength $V_t/(\hbar\omega_c)$.
	     The Fourier transform is taken after
	     the excitation pulse has vanished. $g_\gamma = 0.08$,
	     $N_\mathrm{e} = 1$, $pq = 2$, $T = 1.0$ K, and $L= 100$ nm.
	     $E_\gamma = 1.4$ meV, $\hbar\Gamma = 0.5$ meV,
 	     and $\hbar\omega_\mathrm{ext}=3.5$ meV.}
\label{FP-Vt-Ng}
\end{figure}

\begin{figure}[htb]
    \includegraphics[width=0.48\textwidth]{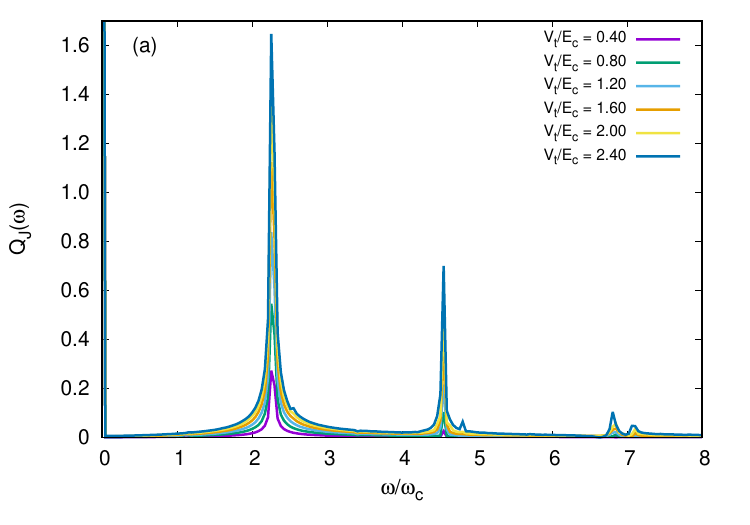}
    \includegraphics[width=0.48\textwidth]{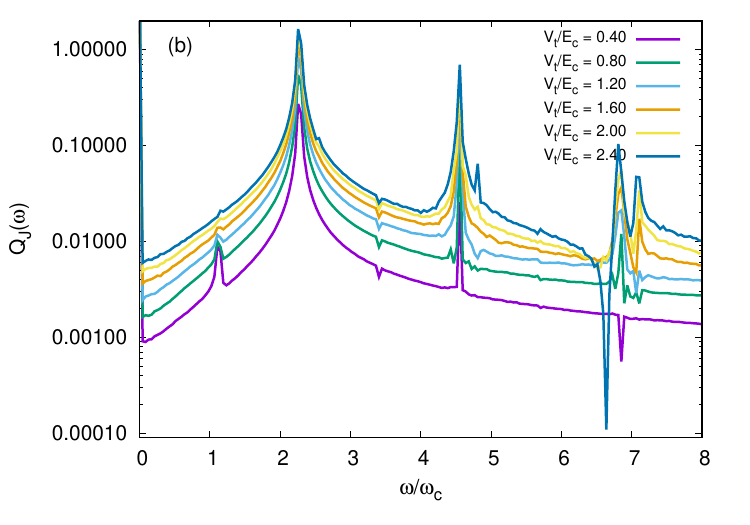}
\caption{The Fourier power spectra for the excitation of the mean orbital
         magnetization $Q_J$ for the system for
	     several values of excitation strength $V_t/(\hbar\omega_c)$ with a linear (a),
	     or logarithmic scale (b). The Fourier transform is taken after
	     the excitation pulse has vanished. $g_\gamma = 0.08$,
	     $N_\mathrm{e} = 1$, $pq = 2$, $T = 1.0$ K, and $L= 100$ nm.
	     $E_\gamma = 1.4$ meV, $\hbar\Gamma = 0.5$ meV,
	     and $\hbar\omega_\mathrm{ext}=3.5$ meV.}
\label{FP-Vt-Qj}
\end{figure}

\begin{figure}[htb]
    \includegraphics[width=0.34\textwidth,bb = 20 20 150 246]{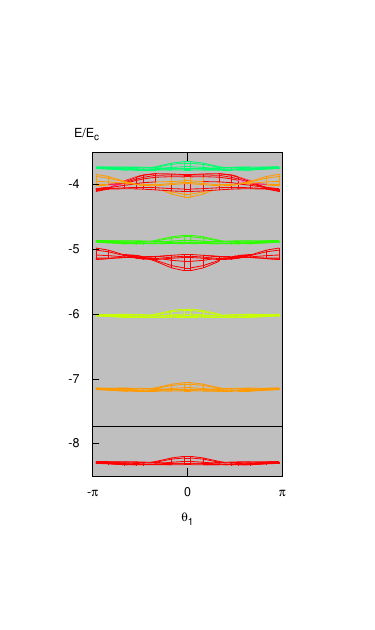}
    \vspace*{-1.5cm}
    \caption{The energy bandstructure of only the lower in energy spin component
            projected on the $\theta_1$ direction in the 1st
            Brillouin zone for $N_\mathrm{e} = 1$, and $pq=2$  The color of the bands indicates their photon content with red
            for 0 and violet for 10.
            The chemical potential $\mu$ is shown by the
            horizontal black line.
            $E_c = \hbar\omega_c = 1.4291$ meV, $g_\gamma = 0.08$,
            $T = 1.0$ K, $L= 100$ nm, and
            $E_\gamma = 1.4$ meV.}
    \label{EH-Vt-Sz1}
\end{figure}

In Subsection \ref{varEg} it became clear that the excitation of the system can take
it outside the regime where linear response is appropriate to describe its time evolution.
This opens the question how the strength of the excitation influences its response if
other parameters are kept constant.
Fig.\ \ref{FP-Vt-Ng} displays the $N_\gamma (\omega)$ excitation peak for different values
of the excitation strength $V_t$. For low values of the excitation strength only one peak
is seen but the height of the peaks seems to grow in a slightly superlinear fashion with
increasing $V_t$, but no shift is seen. Both the mean total energy and the mean photon
number grow in a similar fashion showing that the system is not in the linear response
regime as the excitation strength $V_t$ is increased. For the highest values of $V_t$ a small
side peak appears, that reflects the fact that neither the ground state energy subband nor
the second photon replica of it are totally flat in reciprocal space.
Like in Subsection \ref{varEg} the $Q_J(\omega )$ excitation
shows more structure as
the excitation strength increases as Fig.\ \ref{FP-Vt-Qj} shows.
Again the diamagnetic excitations are stronger than the paramagnetic ones, and the
oscillator strength of the paramagnetic excitations looses relative strength compared
to the diamagnetic ones as $V_t$ increases. For the higher diamagnetic peaks more complex
structure emerges with stronger $V_t$. This modified structure can here be
correlated to changes of high lying subbands caused by their interactions due to
Rabi resonances as is shown in Fig.\ \ref{EH-Vt-Sz1}. The figure shows only the
energy subbands of the lower in energy spin component. Below the chemical
potential is the occupied lowest subband.
Above it are the unoccupied first and second photon replicas of the lowest subband,
respectively with a simple structure. The third and in particular the fourth photon
replica of the ground
subband show complex structures resulting from both the effects of the periodic
superlattice and the electron-photon interaction. The fourth photon replica determines
the excitation peaks seen around $\omega/\omega_c\approx 4.5$.
The change of a photon replica subband in reciprocal space modifies the van Hove singularities
in its density of states with accompanying possible splitting of excitation peaks
\cite{Weiss90:88,PhysRev.89.1189}.
The excitation peaks convey information about the bandstructure of the system.

Complex structures caused by Rabi resonances and the periodicity of the superlattice were seen
relatively lower (compared to the chemical potential) in the energy spectrum for a higher
number of electrons in a dot in Fig.\ 5 in Ref.\ \cite{PhysRevB.109.235306}.

\subsection{One electron in a dot, $N_\mathrm{e}=1$, variation of the
            electron-photon coupling strength $g_\gamma$}
\label{vargg}
In this subsection we explore how the excitation spectra change when the electron-photon
interaction strength $g_\gamma$ is varied. The influence of $g_\gamma$ can be expected to
be complex as it enters the system through two different types of interactions
to different order.

\begin{figure}[htb]
    \includegraphics[width=0.48\textwidth]{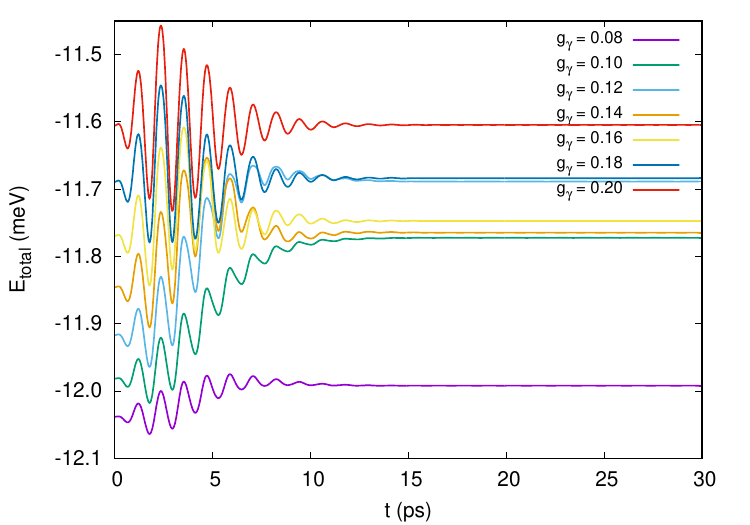}
\caption{The time evolution of the total energy for the first 30 ps for
	     different values of the dimensionless electron-photon coupling $g_\gamma$.
	     $E_\gamma = 1.4$ meV,
	     $N_\mathrm{e} = 1$, $pq = 2$, $T = 1.0$ K, and $L= 100$ nm.
	     $V_t/(\hbar\omega_c) = 0.8$, $\hbar\Gamma = 0.5$ meV,
	     and $\hbar\omega_\mathrm{ext}=3.5$ meV.}
\label{Ne01-Etot-vargg}
\end{figure}

The first 30 ps of the time evolution of the total energy of the system are
seen in Fig.\ \ref{Ne01-Etot-vargg}.
For the lower values of $g_\gamma$ the system is clearly not in the linear
response regime, but for the highest 3 values it is very close to it.
The photon energy is here $E_\gamma = 1.40$ meV and as was shown in
Subsection \ref{varEg} the parameter choice for the excitation pulse lead to finite
energy to be pumped into the system. As the coupling is increased the processes
active in the system are lifted above the energy regime in which the excitation
pulse easily supplies energy to it.

\begin{figure}[htb]
    \includegraphics[width=0.42\textwidth]{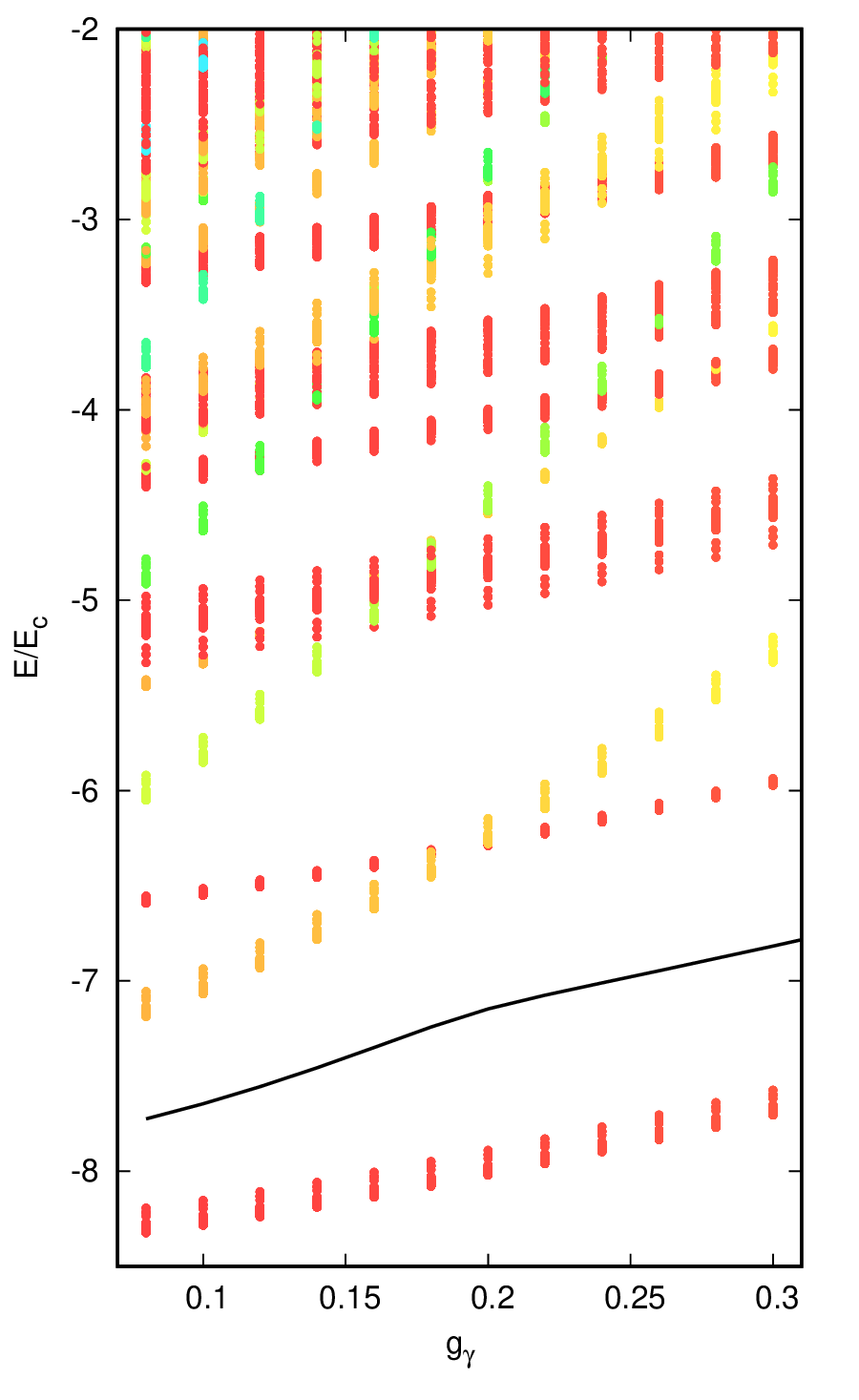}
\caption{The width of the energy subbands as a function of the dimensionless electron-photon
	coupling $g_\gamma$. The photon content of the subbands is
	encoded in the color with red for 0 and violet for 10 photons.
	The black curve represents the chemical potential $\mu$.
	$E_\gamma = 1.4$ meV,
	$pq = 2$ corresponding to $B = 0.8271$ T or $E_c = \hbar\omega_c = 1.4291$ meV,
	$N_\mathrm{e} = 1$, $T = 1.0$ K, and $L= 100$ nm.}
\label{E_gg}
\end{figure}

The static energy spectrum as a function of $g_\gamma$ is displayed in Fig.\ \ref{E_gg}.
We note that none of the subbands is horizontal as even the occupied ground state
acquires a small photon component that grows with increasing coupling.
The electron-photon coupling determines the character of the
emerging quasi-particle states, the cavity-photon dressed electron states.

\begin{figure}[htb]
    \includegraphics[width=0.48\textwidth]{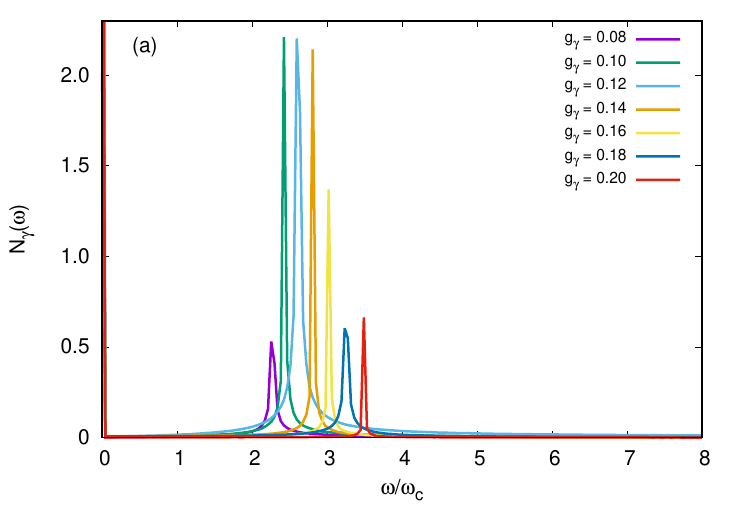}
    \includegraphics[width=0.48\textwidth]{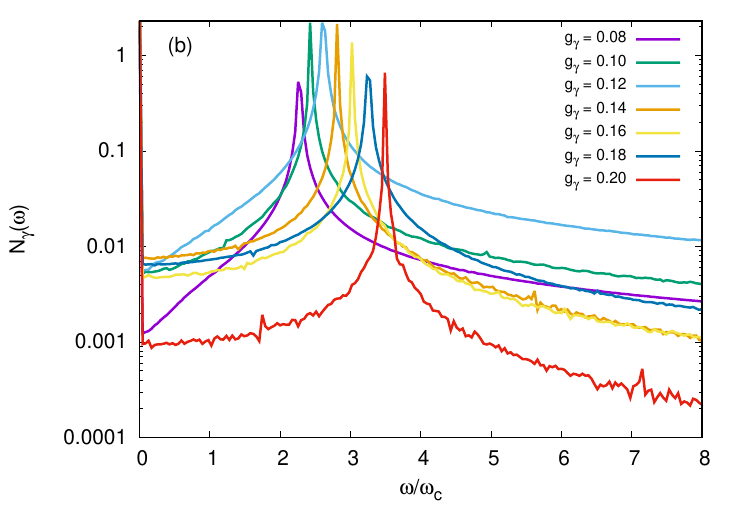}
\caption{The Fourier power spectra for the excitation of the total mean photon number
	     $N_\gamma$ for the system for
	     several values of the dimensionless electron-photon coupling $g_\gamma$
	     with a linear (a),
	     or logarithmic scale (b). The Fourier transform is taken after
	     the excitation pulse has vanished. $E_\gamma = 1.4$ meV,
	     $N_e = 1$, $pq = 2$, $T = 1.0$ K, and $L= 100$ nm.
	     $V_t/(\hbar\omega_c) = 0.8$, $\hbar\Gamma = 0.5$ meV,
	     and $\hbar\omega_\mathrm{ext}=3.5$ meV.}
\label{FP-gg-Ng}
\end{figure}

Fig.\ \ref{FP-gg-Ng} presents the excitation spectra for the different values of
$g_\gamma$ on a linear (a) and a logarithmic (b) scale.
The photon energy is $E_\gamma = 1.4$ meV.
The height of the excitation peaks could not have been easily guessed without
a calculation, as it neither strictly follows the amount of energy pumped into the
system nor a simple function of the dimensionless coupling $g_\gamma$ or the
total energy of the system. The reasons for this have to be partially sought in the
static energy spectrum in Fig.\ \ref*{E_gg}. The structure of the subbands in the energy
spectrum above $E/E_c \approx -5$ is extremely complex regarding crossings,
anticrossings, and photon content, and can not be easily demonstrated by the band
widths. It is even difficult to resolve it clearly with surface plots in reciprocal
space, but this behavior hides Rabi resonances for both parts of the electron-photon
interactions, that strongly modify higher order virtual and real processes.
The location of the peaks, or the distance between them, is not linear with
respect to $g_\gamma$ but clearly demonstrates an additional $g^2_\gamma$-component,
indicating that simple perturbational ideas may not always be sufficient to identify the
nature of the virtual and real processes that contribute to the peaks, though the main
contribution can be clear in some cases.

\begin{figure}[htb]
    \includegraphics[width=0.48\textwidth]{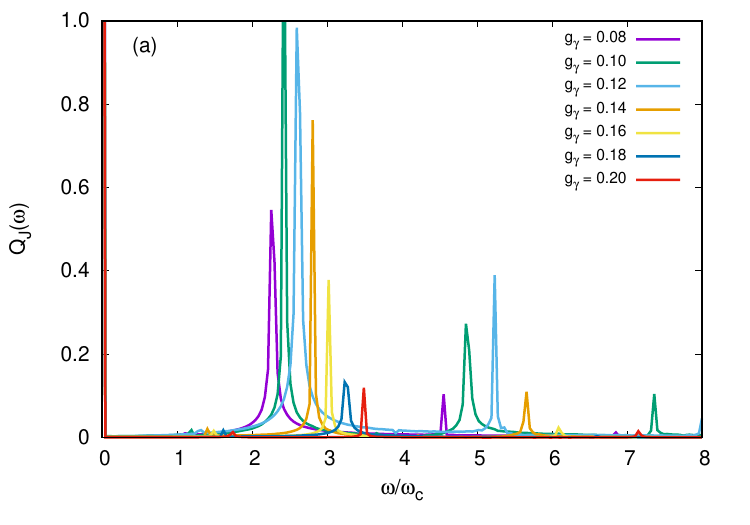}
    \includegraphics[width=0.48\textwidth]{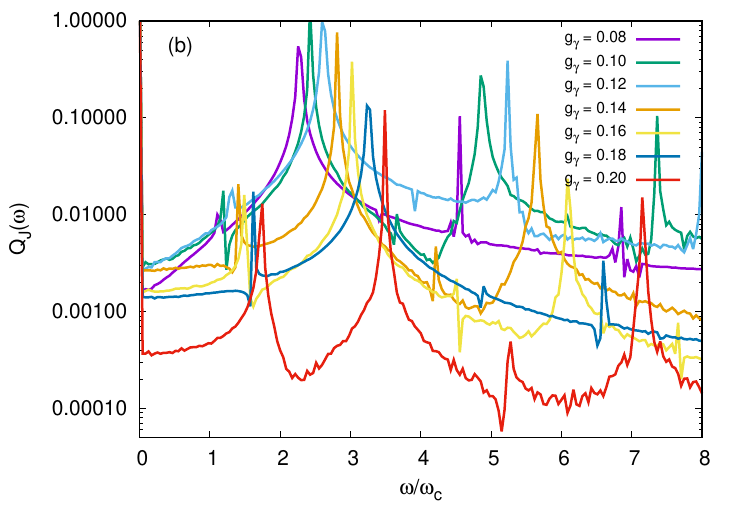}
\caption{The Fourier power spectra for the excitation of the dynamical orbital
	     magnetization $Q_J$ for the system for
	     several values of the dimensionless electron-photon coupling $g_\gamma$
	     with a linear (a),
	     or logarithmic scale (b). The Fourier transform is taken after
	     the excitation pulse has vanished. $E_\gamma = 1.4$ meV,
	     $N_e = 1$, $pq = 2$, $T = 1.0$ K, and $L= 100$ nm.
	     $V_t/(\hbar\omega_c) = 0.8$, $\hbar\Gamma = 0.5$ meV,
	     and $\hbar\omega_\mathrm{ext}=3.5$ meV.}
\label{FP-gg-Qj}
\end{figure}

Further higher order and paramagnetically active excitations are seen in the
excitation spectra for $Q_J(\omega )$ shown in Fig.\ \ref{FP-gg-Qj}.
The second order diamagnetic excitation peaks are slightly spread in energy (or $\omega$)
compared to the first order ones. This reflects properties of the static energy spectrum
in Fig.\ \ref{E_gg}, where it is seen that the higher photon replicas tend to have a higher
slope with respect to $g_\gamma$ than the lower ones.

\subsection{More than 1 electron in a dot, $N_\mathrm{e}>1$}
\label{Ne2-pq1}
In this subsection we explore the excitation of system with 2 and 3 electrons at
the low magnetic field $B = 0.4135$ T corresponding to $E_c = \hbar\omega_c = 0.7145$ meV,
and the number of magnetic flux quanta $pq = 1$ through the unit cell of the
square lattice.
The photon energy is chosen to be $E_\gamma = 1.0$ meV corresponding to the ratio
$E_\gamma /E_c \approx 1.4$.

\begin{figure}[htb]
\includegraphics[width=0.48\textwidth]{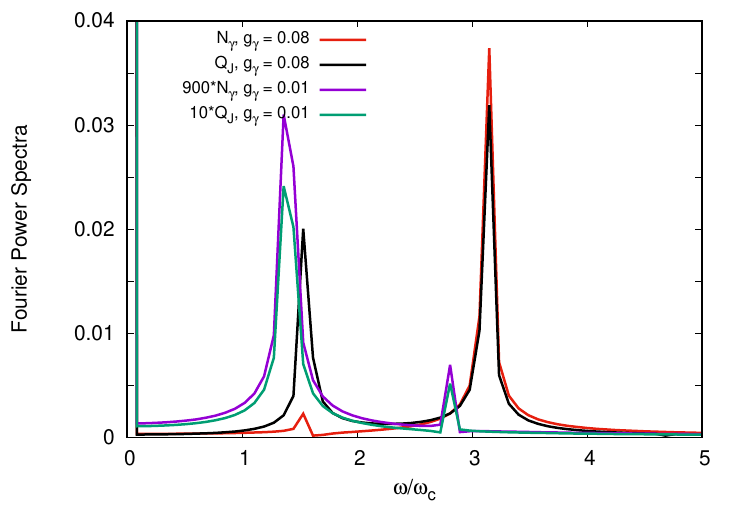}
\caption{The Fourier power spectra for the excitation of the total mean photon number
		 $N_\gamma$ and the dynamic orbital magnetization $Q_J$ for the system for
		 two values of the dimensionless electron-photon coupling $g_\gamma$.
		 The Fourier transform is taken after
		 the excitation pulse has vanished. $E_\gamma = 1.0$ meV,
		 $N_\mathrm{e} = 2$, $pq = 1$, $T = 1.0$ K, and $L= 100$ nm.
		 $V_t/(\hbar\omega_c) = 0.8$, $\hbar\Gamma = 0.5$ meV,
		 $\hbar\omega_\mathrm{ext}=3.5$ meV, and $E_c = \hbar\omega_c = 0.7145$ meV.}
	\label{FP-gg-Ne02}
\end{figure}

The excitation spectrum for $N_\mathrm{e}=2$ is shown in Fig.\ \ref{FP-gg-Ne02}.
In this case no photon replica is occupied initially and the two electrons are in a
singlet spin state without an enhanced spin splitting. For the lower coupling
$g_\gamma = 0.01$ the overlapping peaks close to $\omega/\omega_c = 1.4$ have a
main contribution from a one photon paramagnetic transition. The corresponding peaks
close to $\omega/\omega_c = 2.8$ are much lower and can be assigned to a two photon
diamagnetic transition. The diamagnetic peaks for $g_\gamma = 0.01$ are very small
as $g^2_\gamma << g_\gamma$. This size ratio for the dia- and the paramagnetic excitation
peaks turns around for the strong coupling $g_\gamma = 0.08$ and the peaks get blue-shifted
away from the simple ratios of $E_\gamma /E_c \approx 1.4$ or 2.8 when higher order
interaction effects start to contribute to them as the electron-photon coupling is
increased.

\begin{figure}[htb]
\includegraphics[width=0.48\textwidth]{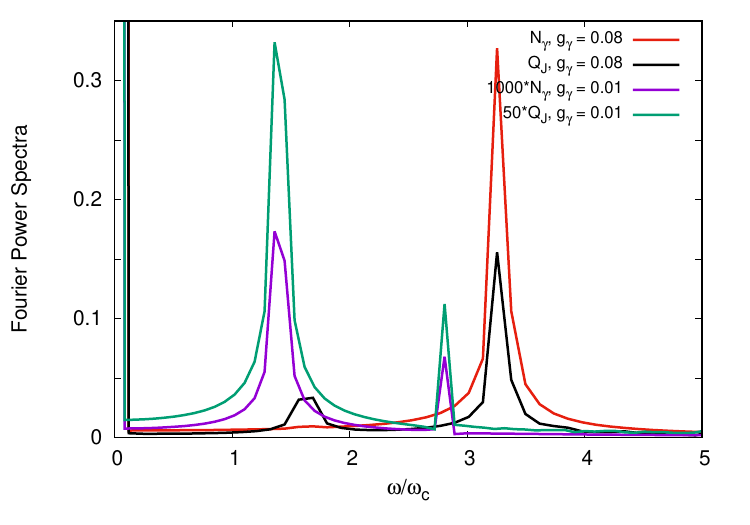}
\caption{The Fourier power spectra for the excitation of the total mean photon number
	     $N_\gamma$ and the dynamic orbital magnetization $Q_J$ for the system for
	     two values of the dimensionless electron-photon coupling $g_\gamma$.
	     The Fourier transform is taken after
	     the excitation pulse has vanished. $E_\gamma = 1.0$ meV,
	     $N_\mathrm{e} = 3$, $pq = 1$, $T = 1.0$ K, and $L= 100$ nm.
	     $V_t/(\hbar\omega_c) = 0.8$, $\hbar\Gamma = 0.5$ meV,
	     $\hbar\omega_\mathrm{ext}=3.5$ meV, and $E_c = \hbar\omega_c = 0.7145$ meV.}
\label{FP-gg-Ne03}
\end{figure}

Similar behavior for the excitation spectra can be seen for the case of
3 electrons in a quantum dot presented in Fig.\ \ref{FP-gg-Ne03}.
Here all parameters are the same as for the case of two electrons in a dot, but
now one photon replica is occupied and the Coulomb exchange interaction leads to
a large spin splitting with approximately one unpaired spin. Like, seen earlier in
two different approaches for the static calculations, the large spin splitting is
slightly reduced by the strong electron-photon interaction for $g_\gamma = 0.08$
\cite{PhysRevB.106.115308,PhysRevB.109.235306}. The electron-photon coupling tends
to reduce the Coulomb exchange interaction.

For both the cases of 2 and 3 electrons in a quantum dot presented here, the excitation
does not add much energy to the systems for the chosen parameters. It remains very
close to the linear response regime.
The increased half-width of the excitation peaks compared to the $pq=2$
results in the previous subsections is due to the smaller cyclotron energy $\hbar\omega_c$
and the increased width of the energy subbands at the lower magnetic field, where the electrons
in the dots are more influenced by electrons in the neighboring dots in the square lattice.

We have performed calculations for a higher number of electrons in a quantum dot.
For a low cavity photon energy resulting in more photon replicas to be occupied initially,
the number of iterations in each time step tends to grow and it can take unreasonable CPU-time
to gather enough time points for accurate Fourier transforms. This reminds us that
the effective electron-photon coupling strength is not only set by the single
parameter $g_\gamma$.

\section{Conclusions}
\label{Conclusions}
We use a DFT-QED-TP approach for the Coulomb interacting
electrons of a 2DEG in a square array of
quantum dots in a homogeneous external perpendicular magnetic field. The 2DEG is placed in a
circular cylindrical FIR photon cavity supporting one TE$_{011}$ mode. The coupling of the 2DEG
to the cavity photons is described by a configuration interaction approach that is updated in
each DFT iteration in both the static case, and in each time step of the dynamical evolution of
the system after a short excitation.

The vector potential for the external magnetic field, and the cavity vector potential in
the long wavelength approximation have the same spatial form.
The electron-photon coupling includes both the para- and the diamagnetic interaction parts,
and the quantized excitation pulse conforms to the symmetry of the cavity and
includes both types of the electron-photon interaction.
These choices for the cavity and the excitation allow
us to tailor the excitation to steer the relative strength of resonances that can either be
assigned to para- or diamagnetic electron-photon coupling. It is, for example, possible to
excite almost exclusively diamagnetic processes in the system.

The cylindrical photon cavity and the form of the excitation promote
magnetic dipole and higher order magnetic processes in the system.
The same is true about the effects of the cylindrical cavity on the
static system. The underlying reason for the effectiveness of this is the
fact that the system is in an external homogeneous magnetic field.

The real-time approach to the excitation allows us to choose parameters to go beyond
a simple linear response regime. Energy can be pumped into the system and nonlinear
higher order processes can be activated leading to higher order harmonics of the fundamental
processes, and blue shifting of resonances due to high order virtual and real processes.

This particular excitation mode of the 2DEG does not lead to center of mass oscillations
of the electron system in each dot like a conventional FIR electrical dipole excitation
is known to do, but is closely related to methods used for measurements of the
cyclotron resonance. Both the original and the generalized Kohn theorem are broken.
The obtained excitation spectra reveal information about the
subband structure in the array of coupled quantum dots with special emphasis on the
photon replica subbands, or generally the photon content of the subbands.

Changes to the structure of individual photon replica subbands due to their interactions
through Rabi resonances and the influences of the square superlattice
lead to changes in the van Hove singularities
in their density of states. The difference in the density of states singularities
of two photon replica subbands with active transitions between them exposes
the singularities as splittings of excitation peaks in the
respective excitation spectra. For the case of one electron in each quantum dot this
happens due to Rabi resonances in high lying subbands that start to contribute to
the excitation process as its strength is increased.
For a higher number of electrons in a dot this regime is expected to happen for a
lower excitation strength.

As the electron number in each dot is increased, weak quadrupole and monopole (breathing)
excitation modes for the charge density can be seen appearing in the time dependent
induced density. The monopole oscillations are connected with the rotational current
oscillation through the Lorentz force, but the quadrupole modes arise due to the square
symmetry of the superlattice.

It is important to be aware of that sometimes
in our analysis we tend to use concepts to explain occurring phenomena
in terms of perturbation theory, even though the results show us that this view has
its limits as we are dealing with a model that is solved iteratively self-consistently
for all the interactions, the Coulomb-interaction, and the electron-photon interaction.
The QED-DFT-TP approach gives us a methodology to access physical phenomena in a
nonperturbative regime of an extended solid state electron systems in a photon cavity.
The purity, or the mobility, and the polarizability of a 2DEG in a GaAs heterostructure
make it an ideal system to explore these phenomena.

\begin{acknowledgments}
This work was financially supported by the Research Fund
of the University of Iceland grant No.\ 92199, and the Icelandic Infrastructure Fund
for ``Icelandic Research e-Infrastructure (IREI)''.
The computations were performed on resources
provided by the Icelandic High Performance Computing
Center at the University of Iceland.
V.\ Mughnetsyan and V.\ Gudmundsson acknowledge support
by the Higher Education and Science Committee of Armenia (grant No.\ 21SCG-1C012).
V.\ Gudmundsson acknowledges support for his visit to the National Taiwan University from the National Science and Technology Council, Taiwan under Grants No.\ NSTC 113-2811-M-002-001
and No.\ NSTC 112-2119-M-002-014.

H.-S.\  Goan acknowledges support from the National Science and Technology Council, Taiwan under Grants No.~NSTC 112-2119-M-002-014, No.\ NSTC 111-2119-M-002-007, and No.\ NSTC 111-2627-M-002-001, and from the National Taiwan University under Grants No.\ NTU-CC-112L893404 and No.\ NTU-CC-113L891604. H.-S.\ Goan is also grateful for the support from the ``Center for Advanced Computing and Imaging in Biomedicine (NTU-112L900702)''  through The Featured Areas Research Center Program within the framework of the Higher Education Sprout Project by the Ministry of Education (MOE), Taiwan, and the support from the Physics Division, National Center for Theoretical Sciences, Taiwan.

J.-D.\ Chai acknowledges support from the National Science and Technology Council,
Taiwan under Grants No.\ NSTC113-2112-M-002-032 and No.\ MOST110-2112-M-002-045-MY3.
J.-D.\ Chai is also grateful for the support from the Physics Division, National Center
for Theoretical Sciences, Taiwan.

C.-S.\ Tang acknowledges funding support by the National United
University through Contract No.\ 113-NUUPRJ-01.

V.\ Moldoveanu acknowledges financial
support from the Core Program of the National Institute of Materials Physics, granted by the Romanian Ministry
of Research, Innovation and Digitalization under the Project PC2-PN23080202.

\end{acknowledgments}

%
\appendix

%


%

\end{document}